\documentclass{ws-p8-50x6-00-mod}

\def \sss  {\scriptscriptstyle}
\def \pt   {p_{\rm\sss T}}
\def \ptq  {p_{\rm\sss T}^2}
\def \as   {\alpha_{\sss S}}
\def \ep   {\epsilon}
\newcommand\aem{\alpha_{\rm em}}

\def    \jhep   #1#2#3{{\it JHEP} {\bf #1}(#2)#3}
\def    \np     #1#2#3{{\it Nucl. Phys.} {\bf #1}(#2)#3}

\def    \epj    #1#2#3{{\it Eur. Phys. J.} {\bf #1}(#2)#3}

\def    \hepph  #1 {{\tt hep-ph/#1}}
\def    \hepex  #1 {{\tt hep-ex/#1}}

\begin{document}

\title{The $\pt$ spectrum of heavy quarks in photoproduction}

\author{Stefano Frixione}

\address{LAPP, Annecy, France and INFN, Sezione di Genova, Italy}

\author{Matteo Cacciari}

\address{Universit\`a di Parma, Italy}

\author{Paolo Nason}

\address{INFN, Sezione di Milano, Italy}

\maketitle

\abstracts{
We illustrate a formalism that allows to match the next-to-leading
order cross section for the photoproduction of heavy quarks to the cross
section obtained by resumming logarithms of $p_{\sss T}/m$ to the
next-to-leading accuracy, thus giving a sensible prediction for any 
value of $\pt$. We present a comparison between our predictions and 
H1 and ZEUS data.
}

\section{Introduction}

The computations of the cross sections for the production of open
heavy flavours at fixed-order (FO) in perturbative QCD have been
so far carried out at the next-to-leading order (NLO) accuracy
for all possible production mechanisms. For single-inclusive 
observables, when the transverse momentum $\pt$ of the quark 
is much larger than its mass $m$, terms containing $\log(\pt /m)$
become large in the FO cross section, which is thus no longer reliable. 
In such a case, techniques are used that allow the resummation
to all orders of the leading (LL) and next-to-leading (NLL) towers 
of logarithms of $\pt /m$; in order to achieve the resummation, the 
heavy quark is effectively treated as massless.

The resummed (RS) and FO cross sections are therefore relevant to
complementary regions in $\pt$. The problem then arises whether to
compare the data to FO or to RS predictions, since the inequality
$\pt\gg m$ cannot be unambiguously translated into a quantitative statement.
It is thus desirable to write the single-inclusive cross section in a form
that is sensible in the whole $\pt$ range, that is, which interpolates
between the FO result, relevant to the small- and intermediate-$\pt$
regions, and the RS result, relevant to the large-$\pt$ region.
This is the aim of ref.~\cite{CGN} and ref.~\cite{CFN}, relevant to
hadro- and photoproduction respectively. In this letter, we shall 
concentrate on the latter, reviewing the formalism and presenting 
phenomenological predictions for charm production at HERA.

Our master formula reads:
\begin{equation}
  \label{eq:merge}
  \mbox{FONLL}=\mbox{FO}\;+\left( \mbox{RS}\; -\; \mbox{FOM0}\right)\;
\times G(m,\pt)\;,
\end{equation}
where FONLL (for Fixed Order plus Next-to-Leading Logarithms) will give
sensible predictions in the whole $\pt$ range, and FOM0 is obtained
from FO by letting to zero all the terms suppressed by powers of
$m/\pt$. The subtraction of FOM0 from RS in eq.~(\ref{eq:merge})
is necessary to avoid double counting, since some of the logarithms
appearing in RS are already present in FO. To be more precise, FONLL
will have the following features:
\begin{itemize}
\vspace*{-0.5pc}
\item
All terms of order $\aem\as$ and $\aem\as^2$ are included exactly,
including mass effects;
\vspace*{-0.5pc}
\item
All terms of order $\aem\as\left(\as\log\pt/m\right)^i$
and $\aem\as^2\left(\as\log\pt/m\right)^i$
are included, with the possible exception of
terms that are suppressed by powers of $m/\pt$.
\end{itemize}
\vspace*{-0.5pc}
Finally, the function $G(m,\pt)$ is rather arbitrary, except that it must
be a smooth function, and that it must approach one when $m/\pt\,\to\,0$, 
up to terms suppressed by powers of $m/\pt$.  In what follows, we shall use
\begin{equation}
G(m,\pt)=\frac{\pt^2}{\pt^2 + c^2 m^2}\;,
\end{equation}
with $c$ a free parameter. The practical implementation of eq.~(\ref{eq:merge})
is rather involved, especially in the photoproduction case; all the details
can be found in ref.~\cite{CFN}

\section{Charm production at HERA}

In this section, we shall compare the predictions based upon
eq.~(\ref{eq:merge}) to the experimental data for $D^*$ meson
photoproduction obtained at the HERA collider by the H1~\cite{H1paper}
and the ZEUS~\cite{ZEUSpaper} collaborations. We set $m$=1.5 GeV, and
the renormalization and factorization scales equal to the transverse
mass of the quark, $\sqrt{\ptq+m^2}$. The parton densities in the
proton (photon) are given by the CTEQ5M1 (AFG) set. We set
$\Lambda_{\sss QCD}^{(5)}=226$~MeV, as constrained by the CTEQ5M1 
set; this value is almost identical to the central value of the 
PDG global fit. Finally, in order to simulate the hadronization 
of the bare $c$ quarks into $c$-flavoured mesons, we use the
Peterson function, and multiply the results by the probability
of a $c$ quark fragmenting into a $D^*$ meson, $P_{c\to D^*}$.
In order to match the experimental conditions, we also need to
apply cuts on the variables $y$ and $Q^2$, that enter the Weizs\"acker-%
Williams function (the latter variable is the upper limit of the photon
virtuality squared). In particular, in the case of the ZEUS data,
we have:
\begin{equation}
0.187<y<0.869,\;\;\;\;Q^2=1~{\rm GeV}^2.
\end{equation}
H1 has two data samples, corresponding to the two electron taggers
(ETAG33 and ETAG44) used in the analyses:
\begin{eqnarray}
0.29<y<0.62,\;\;\;\;Q^2=0.01~{\rm GeV}^2:\;\;\;\;\;\;{\rm ETAG33};\\
0.02<y<0.32,\;\;\;\;Q^2=0.009~{\rm GeV}^2:\;\;\;\;\;\;{\rm ETAG44}.
\end{eqnarray}
We also point out that ZEUS use $P_{c\to D^*}=22.2\%$, and H1 use
$P_{c\to D^*}=27\%$.

\begin{figure}[H]
  \begin{center}
    \epsfig{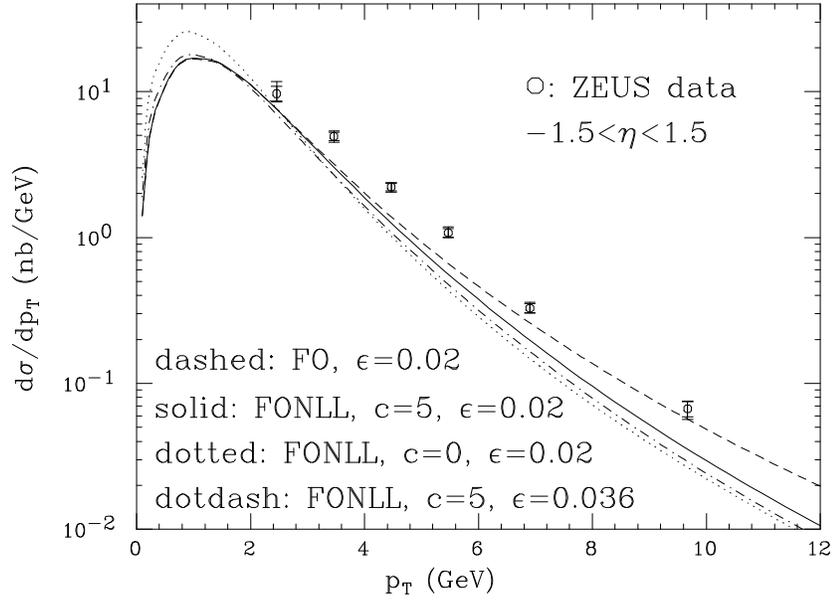}
\caption{\label{fig:zeus-pt} 
$\pt$ spectrum of $D^*$ mesons, in the visible region of the ZEUS
experiment. ZEUS data are compared to the theoretical predictions.
See the text for details.
}
  \end{center}
\end{figure}
\begin{figure}[H]
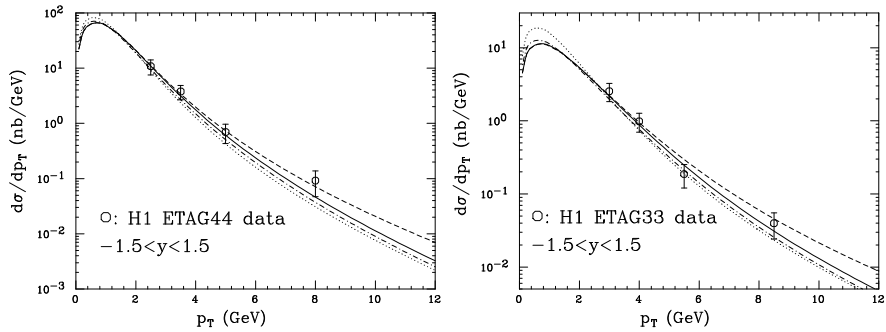

  \begin{center}
    \epsfig{figure=h1_pt_tag44.ps,width=13.7pc}
    \epsfig{figure=h1_pt_tag33.ps,width=13.7pc}
\caption{\label{fig:h1-pt} 
$\pt$ spectrum of $D^*$ mesons, in the visible regions of the H1
experiment: ETAG44 (left panel), and ETAG33 (right panel).
H1 data are compared to the theoretical predictions. The 
patterns of the theoretical predictions are as in fig.~\ref{fig:zeus-pt}.
}
  \end{center}
\end{figure}

In figure~\ref{fig:zeus-pt}, ZEUS data are compared to our theoretical
predictions. The solid line is obtained from eq.~(\ref{eq:merge}),
by setting $c=5$ and the Peterson parameter $\ep=0.02$. The dotted
(dot-dashed) line is obtained with $c=0$ and $\ep=0.02$ ($c=5$ and
$\ep=0.036$). Finally, the dashed line is the FO result (at NLO
accuracy), with $\ep=0.02$. It appears that none of the theoretical
curves describes the experimental data particularly well, both in
normalization and in shape.
The impact of unknown power-suppressed terms, and of terms beyond 
NLO in the perturbative expansion of the resummed cross sections, 
can be roughly estimated by looking at the difference between the 
matched results obtained with $c=5$ and $c=0$ (solid and dotted 
curves respectively). This difference is non negligible, but it does 
not help in understanding the discrepancy with the data. On the other
hand, the comparison between theory and H1 data, fig.~\ref{fig:h1-pt},
appears to be satisfactory. The same pattern is found in the rapidity
and pseudorapidity spectra: H1 data are well reproduced by QCD,
ZEUS data are not. A more detailed comparison between theory
and HERA data will be presented in a forthcoming publication, where
a complete study of the dependence upon the parameters entering the
calculation (such as the mass and the scales) will be carried out.

One of the authors (S.F.) wishes to thank C. Grab and L. Gladilin
for useful discussions on H1 and ZEUS experiments, and CERN TH
Division for the hospitality during the course of this work.

\end{document}